\documentclass[a4paper,fleqn]{cas-dc}

\usepackage[numbers]{natbib}
\usepackage[utf8]{inputenc}
\usepackage{subfig}
\usepackage{hyperref}

\def\tsc#1{\csdef{#1}{\textsc{\lowercase{#1}}\xspace}}
\tsc{WGM}
\tsc{QE}
\tsc{EP}
\tsc{PMS}
\tsc{BEC}
\tsc{DE}

\begin{document}
\let\WriteBookmarks\relax
\def\floatpagepagefraction{1}
\def\textpagefraction{.001}
\shorttitle{Linear cosmological Perturbations ...}
\shortauthors{someone et~al.}

\title [mode = title]{Linear cosmological Perturbations in scalar-tensor-vector gravity}                      



\author[1]{Sara Jamali}
\ead{sara.jamali@um.ac.ir}


\address[1]{Department of Physics, Faculty of Science, Ferdowsi University of Mashhad
P.O. Box 1436, Mashhad, Iran}

\author[1,2]{Mahmood Roshan} 
\cormark[1]
\ead{mroshan@um.ac.ir}



\address[2]{School of Astronomy, Institute for Research in Fundamental Sciences (IPM), P.O. Box 19395-5531, Tehran, Iran}

\author[3]{Luca Amendola} 
\ead{l.amendola@thphys.uni-heidelberg.de}

\address[3]{Institute for Theoretical Physics, University of Heidelberg,  Philosophenweg 16, D-69120 Heidelberg, Germany}

\cortext[cor1]{Corresponding author}


\begin{abstract}
We investigate the cosmological perturbations in the context of a Scalar-Tensor-Vector theory of Gravity known as MOG in the literature. Recent investigations show that MOG reproduces a viable background cosmological evolution comparable to $\Lambda$CDM. However, the matter dominated era is slightly different. In this paper, we study the  linear matter perturbations and estimate the relevant modified gravity parameters. We show that MOG reduces the growth rate of the perturbations and comparing with the RSD data reveals that MOG suggests a higher value for $\sigma_8$, compare to $\Lambda$CDM. This point, constitute a powerful challenge to the cosmological viability of MOG.

\end{abstract}




\maketitle

\section{Introduction}

The Scalar-Vector-Tensor theory of gravity, also known as MOG in the
literature, has been introduced in \cite{Moffat:2005si}.
MOG does not includes dark matter but introduces instead two scalar fields, $G$ and $\mu$, and one vector
field, $\phi_{\alpha}$, in addition to the metric tensor. Although MOG is plagued by ghosts, it does not suffer from the tachyonic instability, which means that in the non-quantum limit, that is suitable for the cosmological implications, MOG can be considered as a viable theory \cite{Jamali:2017zrh}. We assume the quantum ghost instability can be cured by additional terms relevant at high energies but unimportant at the classical low-energy level investigated in this paper. Although developing a ghost free version of MOG would be interesting, in order to compare our results with those already claimed in the literature, we use the original version of MOG investigated almost in all the previous works.

The astrophysical consequences of MOG have been widely investigated in,  for example, \cite{Moffat:2013uaa}-\cite{Israel:2016qsf}. In \cite{Moffat:2013uaa}, it has been shown that MOG can explain the flat rotation curve of spiral galaxies without invoking dark matter particles. It is shown in \cite{Moffat:2013sja} that MOG can also explain the mass discrepancy in galaxy clusters.
 In the strong field limit, MOG black holes have been also investigated in the literature, for example see \cite{Sheoran:2017dwb}-\cite{Armengol:2016xhu}. While MOG is consistent with the recently discovered gravitational wave signals \cite{Green:2017qcv}, it has been shown in \cite{Manfredi:2017xcv} that the quasinormal modes of gravitational perturbations in the ringdown phase of the merging of two MOG black holes have different frequencies compared to those of GR. In \cite{Jamali:2016zww}, using dynamical system approach, It is shown that The cosmic evolution
 starts from a standard radiation dominated era, evolves
towards a matter dominated epoch and tends to a late time accelerated
phase. In \cite{Jamali:2017zrh}, we showed that MOG cannot fit the observational
data of the sound horizon angular size. However, a slightly modified version of MOG, called mMOG, gives a good agreement with the sound horizon data, although the matter
dominated era of mMOG remains  slightly different from $\Lambda$CDM. 

 It is claimed in the literature, \cite{Shojai:2017tzv,Moffat:2011rp}, that MOG increases the growth rate
of matter perturbations, compared to $\Lambda$CDM. Although they use the modified Poisson equation in MOG obtained  in a non-expanding universe and also ignore the evolution of perturbations in the scalar field $G$.
In this paper we revisit the
linear perturbations in MOG and compare it with $\Lambda$CDM, without any of these  restrictive assumptions. We find a slower growth and comparing MOG to the available redshift space distortion (RSD) data, a higher value for $\sigma_{8}$. Although \emph{per se} this fact does not
rule out MOG, such a high value will probably be in conflict with
lensing and CMB results.

\section{The Scalar-Tensor-Vector theory}
\label{sec-1}

The action of Scalar-Tensor-Vector theory of
Gravity, known also as MOG \cite{Moffat:2007nj} , is
\begin{equation}
S=S_{\text{gravity}}+S_{\text{scalar fields}}+S_{\text{vector field}}+S_{\text{m}}\label{action}
\end{equation}
in which, 
\begin{equation}
S_{\text{gravity}}=\!\int\!\!\sqrt{-g}\ d^{4}x\Big(\frac{R-2\Lambda}{16\pi G}\Big)\label{gr}
\end{equation}
\begin{small}
\begin{equation}
S_{\text{scalar fields}}=\int\sqrt{-g}\ d^{4}x\frac{1}{2 G}g^{\mu\nu}\Big(\frac{\nabla_{\mu} G \nabla_{\nu} G}{G^{2}}+\frac{\nabla_{\mu}\mu\nabla_{\nu}\mu}{\mu^{2}}\Big)\label{scalar}
\end{equation}
\end{small}
\begin{equation}
S_{\text{vector field}}=\int\sqrt{-g}\ d^{4}x\frac{1}{4\pi}\Big(\frac{1}{4}B_{\mu\nu}B^{\mu\nu}+V_{\phi}\Big)\label{vec}
\end{equation}
where $R$ is the Ricci scalar, $\Lambda$ is the effective cosmological
constant and the anti-symmetric tensor $B_{\mu\nu}$ is written as
$\nabla_{\mu}\phi_{\nu}-\nabla_{\nu}\phi_{\mu}$, where $\phi_{\mu}$
is the vector field.

The vector field
potential $V_{\phi}$ is set to $-\frac{1}{2}\mu^{2}\phi_{\alpha}\phi^{\alpha}$,
in which $\mu$, in general, is a scalar field which plays the role of the mass
of the vector field. This potential is the original form introduced
in \cite{Moffat:2005si}, that also leads to a viable weak field limit and
also to an acceptable sequence of cosmic epochs. The action of matter,
$S_{\text{m}}(g_{\alpha\beta},\phi_{\alpha})$, is postulated to be
coupled to the vector field. In this case, there will be a non-zero
fifth force current $J_{\alpha}$. For the sake of simplicity and without loss of generality, we set the scalar field $\mu$
 constant during the structure formation era. It should be mentioned
that this scalar field does not play a crucial role in the cosmic
history of MOG \cite{Moffat:2014bfa} and \cite{Moffat:2015bdaf}. More specifically
it is shown in \cite{Jamali:2017zrh} and \cite{Jamali:2016zww} that
$\mu$ does not carry a substantial contribution to the total energy budget. In contrast, the scalar field $G$
seriously influences the dynamics of the gravitating systems \cite{Moffat:2013sja,Moffat:2013uaa}.

The energy-momentum tensors associated with the scalar field $\mathcal{G}$, defined as  $\mathcal{G}=1/G$, and with
the vector field $\phi_{\alpha}$, are defined as $-\frac{2}{\sqrt{-g}}\frac{\delta S_{\text{field}}}{\delta g^{\mu\nu}}$, and results in
\begin{small}
\begin{align}
&\hspace{-0.6cm}T_{\mu\nu(\mathcal{G})}=-\frac{\nabla_{\mu}\mathcal{G}\nabla_{\nu}\mathcal{G}}{\mathcal{G}}+\frac{1}{2}g_{\mu\nu}\frac{\nabla_{\alpha}\mathcal{G}\nabla^{\alpha}\mathcal{G}}{2\mathcal{G}},
\label{temgi}
\\
&\hspace{-0.6cm}T_{\mu\nu(\phi_{\alpha})}=-\frac{1}{4\pi}\Big[B_{\mu}^{\alpha}B_{\nu\alpha}-g_{\mu\nu}(\frac{1}{4}B^{\rho\sigma}B_{\rho\sigma}+V_{\phi})+2\frac{\partial V_{\phi}}{\partial g^{\mu\nu}}\Big].
\label{tembi}
\end{align}
\end{small}
Now let us briefly
review the field equations of the theory. Variation of the action
\eqref{action} with respect to $g^{\mu\nu}$, yields the following
modified Einstein equation 
\begin{small}
\begin{equation}
\hspace{-0.3cm} G_{\mu\nu} -  \frac{\nabla_{\mu}\nabla_{\nu}\mathcal{G}}{\mathcal{G}}+ g_{\mu\nu}\frac{\Box \mathcal{G}}{\mathcal{G}}+\Lambda \mathcal{G} g_{\mu\nu}
\!=\!\frac{8\pi}{\mathcal{G}}\left(T_{\mu\nu(\mathcal{G})}\!+\!T_{\mu\nu(\phi_{\alpha})}\!+\!T_{\mu\nu(\text{m})}\right)\label{ein}
\end{equation}
\end{small}
where $T_{\mu\nu\text{(m)}}=-\frac{2}{\sqrt{-g}}\frac{\delta S_{\text{m}}}{\delta g^{\mu\nu}}$
and $G_{\mu\nu}$ is the Einstein tensor. On the other hand, by varying the actions $\eqref{scalar}$
and $\eqref{vec}$ with respect to $\mathcal{G}$ and $\phi_{\alpha}$,  the following
field equations can be derived 
\begin{eqnarray}
&&\Box \mathcal{G}=\frac{1}{16\pi}R \mathcal{G} + \frac{1}{2 \mathcal{G}}\nabla_{\alpha}\mathcal{G}\nabla^{\alpha}\mathcal{G}-\frac{\Lambda \mathcal{G}}{8\pi}\label{boxg}
\\
&&\nabla_{\beta}B^{\beta\alpha}=4\pi J^{\alpha}-\mu^{2}\phi^{\alpha}\label{eoms}
\end{eqnarray}
where the d'Alembertian operator $\Box \mathcal{G}$ is defined as $\nabla_{\alpha}\nabla^{\alpha}\mathcal{G}$,
and the fifth force current is obtained by varying the matter action
with respect to the vector field as $J^{\alpha}=\frac{1}{\sqrt{-g}}\frac{\partial S_{\text{m}}}{\partial\phi_{\alpha}}$.
 Now, using \eqref{boxg}, we can find the
following continuity equations for $\mathcal{G}$ and $\phi_{\alpha}$ 
\begin{small}
\begin{align}
& \hspace{-0.8cm} \nabla_{\mu}T_{\nu\,(\phi_{\alpha})}^{\mu}\!=\!B_{\alpha\nu}J^{\alpha}\!-\!\frac{1}{4\pi}\frac{\partial V_{\phi}}{\partial\phi_{\alpha}}B_{\nu\alpha}\!+\!\frac{\nabla_{\nu}V_{\phi}}{4\pi}\!-\!\frac{1}{2\pi}\nabla^{\mu}(\frac{\partial V_{\phi}}{\partial g^{\mu\nu}})\label{vcoupling}
\\
&\hspace{-0.8cm} \nabla_{\mu}T_{\nu\,(\mathcal{G})}^{\mu}=-\frac{R}{16\pi}\nabla_{\nu}\mathcal{G}+\frac{\Lambda}{8\pi}\nabla_{\nu}\mathcal{G}.
\label{Coupling}
\end{align}
\end{small}
We suppose that the matter content of the universe is a perfect fluid. In this case, by assuming the continuity relation $\nabla_{\alpha}J^{\alpha}=0$ (or equivalently $\nabla_{\alpha}\phi^{\alpha}=0$), one finds 
\begin{eqnarray}
\begin{split} &  & \nabla_{\alpha}T_{\nu\,\text{(m)}}^{\alpha}=-B_{\alpha\nu}J^{\alpha},\,\,\,~~~\end{split}
\label{mcoupling1}
\end{eqnarray}
(see \cite{Roshan:2012qy} for more details).
The assumption of isotropy and homogeneity
 leads to $B_{\alpha\nu}=0$ at the background level.
Therefore, by using eq. \eqref{mcoupling1}, one recovers the normal continuity equation for the ordinary matter.

It is convenient now to rewrite the relevant background equations using  the $e$-folding time $\tau=\text{ln}a$ and $\mathcal{H}=Ha$, the conformal Hubble function:
\begin{align}
&\hspace{-0.8cm}\frac{\mathcal{G}''}{\mathcal{G}}=-\frac{\mathcal{G}'\mathcal{H}'}{\mathcal{G}\mathcal{H}}+\frac{\mathcal{G}'^{2}}{2\mathcal{G}^{2}}-\frac{2\mathcal{G}'}{\mathcal{G}}-\frac{3\mathcal{H}'}{8\pi\mathcal{H}}-\frac{3}{8\pi}+\frac{e^{2\tau}\Lambda}{8\pi\mathcal{H}^{2}},\label{bakg}
\\
&\hspace{-0.8cm} J_{0\tau}(\tau)=\frac{\mu^{2}\phi_{0\tau}}{4\pi}\label{bakv}
\end{align}
where a prime stands for derivative with respect to $\tau$. As we
shall see, equations \eqref{bakg} and \eqref{bakv} are necessary
to simplify the first order equations. In the following section, we linearize these equations and investigate the growth of
density perturbations.

\section{The perturbed equations}

\label{sec-2} Let us start with the following perturbed flat-space metric in the Newtonian gauge
\begin{equation}
ds^{2}=e^{2\tau}\bigg[-\left(1+2\Psi\right)\mathcal{H}^{-2}d\tau^{2}+\left(1+2\Phi\right)\delta_{ij}dx^{i}dx^{j}\bigg]\label{pmetric}
\end{equation}
We work from now on in Fourier space, with $\mathbf{k}$ denoting the wavevector.
One can write the perturbed fields $\phi_{\alpha}$ and $\mathcal{G}$
 as 
\begin{align}
 &\hspace{-0.8cm} \phi_{\alpha}=(\phi_{\tau},\phi_{\mathbf{i}})=(\phi_{0\tau}(\tau)+\phi_{1\tau}(\tau)e^{i\mathbf{k}\cdot\mathbf{r}}~,~ik\phi_{1\mathbf{i}}(\tau)e^{i\mathbf{k}\cdot\mathbf{r}}~)\nonumber\\
 &\hspace{-0.8cm} \mathcal{G}(\tau)=\mathcal{G}_{0}(\tau)+\mathcal{G}_{1}(\tau)e^{i\mathbf{k}\cdot\mathbf{r}}
\label{perfields}
\end{align}
where the subscript ${\mathbf{i}}$ stands for any of three spatial
components, i.e. $(x,y,z)$, and both background and perturbed quantities
are functions of $\tau$. From now on, the subscripts $0$ and $1$
specify the background and the first order perturbed fields, respectively.

We  perturb now the energy-momentum tensors. Let us start with the energy-momentum tensor of the ordinary matter. In this case it is straightforward to show that
\begin{align}
& T_{0~(\text{m})}^{0} =  -\left(\rho+\delta~\rho~e^{i\mathbf{k}\cdot\mathbf{r}}\right)\nonumber\\
& T_{0~(\text{m})}^{\mathbf{i}} = \frac{i}{\sqrt{3}k}\rho~\theta\left(\omega+1\right)~e^{i\mathbf{k}\cdot\mathbf{r}}\nonumber\\
& T_{\mathbf{i}~(\text{m})}^{\mathbf{j}} = (\rho~\omega+c_{s}^{2}\delta~\rho~e^{i\mathbf{k}\cdot\mathbf{r}})+\Sigma_{\mathbf{i}}^{\mathbf{j}}\label{tem}
\end{align}
where $\omega$ is the equation of state parameter, $\delta=\frac{\delta\rho}{\rho}$
is the density contrast, $\rho$ is the background density, $\theta=i\mathbf{k}\cdot\mathbf{v}/\mathcal{H}$
is the velocity divergence and $\mathbf{v}$ is the peculiar velocity.
Perturbations in the fluid pressure $p$ is given by $\delta p=c_{s}^{2}\delta\rho$,
where $c_{s}^{2}$ is the adiabatic sound speed of the fluid. Since matter is supposed to be approximated by a
perfect fluid, we ignore the anisotropic stress tensor $\Sigma_{\mathbf{i}}^{\mathbf{j}}$.

Similarly, in the following we linearize the energy-momentum tensors associated with the fields $\mathcal{G}$ and $\phi_{\alpha}$. To do so, we use equation
\eqref{temgi}, and find the first order perturbation of $T_{(\mathcal{G})}^{\mu\nu}$
shown as $\delta T_{(\mathcal{G})}^{\mu\nu}$. The result is
\begin{align}
 &\hspace{-0.6cm} \delta T_{0~(\mathcal{G})}^{0} =\frac{\mathcal{H}^{2}\mathcal{G}_{0}'}{2\mathcal{G}_{0}^{2}}\Big(2\mathcal{G}_{0}\left(\mathcal{G}_{1}'-\Psi \mathcal{G}_{0}'\right)-\mathcal{G}_{1}\mathcal{G}_{0}'\Big)e^{-2\tau+i\mathbf{k}\cdot\mathbf{r}},\nonumber \\
&\hspace{-0.6cm} \delta T_{0~(\mathcal{G})}^{\mathbf{i}} = -\frac{ik\mathcal{G}_{1}\mathcal{G}_{0}'}{\sqrt{3}\mathcal{G}_{0}}e^{-2\tau+i\mathbf{k}\cdot\mathbf{r}}
\label{temgimix}
\end{align}
and $\delta T_{0~(\mathcal{G})}^{0} = -\delta T_{\mathbf{i}~(\mathcal{G})}^{\mathbf{i}}$. In a similar way, for the vector field $\phi_{\alpha}$, using equation
\eqref{tembi}, we have 
\begin{align}
& \hspace{-0.7cm} \delta T_{0~(\phi_{\alpha})}^{0} = -\delta T_{\mathbf{i}~(\phi_{\alpha})}^{\mathbf{i}}=\frac{\mu^{2}\mathcal{H}^{2}\phi_{0\tau}}{4\pi}\Big(\phi_{1\tau}-\phi_{0\tau}\Psi\Big)e^{-2\tau+i\mathbf{k}\cdot\mathbf{r}},\nonumber \\
& \hspace{-0.7cm} \delta T_{0~(\phi_{\alpha})}^{\mathbf{i}} = \frac{ik}{4\pi}\mu^{2}\phi_{1\mathbf{i}}\phi_{0\tau}e^{-2\tau+i\mathbf{k}\cdot\mathbf{r}}
\label{tembimix}
\end{align}
Now, to find the linearized form of the conservation equations
\eqref{vcoupling} and \eqref{Coupling}, we first start
with the scalar field $\mathcal{G}$ and use equations \eqref{Coupling} and
\eqref{temgi} to find the first order relations. In this
case the covariant derivative of \eqref{temgi} and the right hand side of equation \eqref{Coupling} at the perturbed level
can be straightforwardly calculated. Since they are long to be written here, we refer the reader to this explanation, if it is needed. In fact, the spatial component gives rise to a trivial relation. Notice that
to show this, one needs to insert the background equation for $\mathcal{G}_{0}''$
given in \eqref{bakg}. On the other hand, the time component leads
to a second-order differential equation for $\mathcal{G}_{1}$, see equation \eqref{approg}. We will discuss this relation in the next section.

Now we return to the vector field's conservation equation \eqref{vcoupling}.
Let us first use equation \eqref{tembimix} and linearize the left
hand side of equation \eqref{vcoupling}. The result is 
\begin{align}
 &\hspace{-0.8cm} \nabla_{\mu}\delta T_{\text{i}}^{\mu}(\phi_{\alpha})=-\frac{ik\mu^{2}\mathcal{H}^{2}}{12\pi}\phi_{0\tau}\left(\sqrt{3}\phi_{1\tau}-3\phi_{1\text{i}}'\right)e^{-2\tau+i\mathbf{k}\cdot\mathbf{r}}\nonumber\\
 &\hspace{-0.8cm} \nabla_{\mu}\delta T_{0}^{\mu}(\phi_{\alpha})=0\label{emcvector}
\end{align}
By keeping the first order terms on the right-hand side of \eqref{vcoupling},
one can easily show that the time component vanishes. For the spatial components, it turns out that only the first
term on the right-hand side contributes. Therefore the spatial component on the right
hand side of \eqref{vcoupling} in the linear limit is written as
\begin{eqnarray}
(B_{\alpha\text{i}}J^{\alpha})_{1}=-\frac{1}{3}ik\mathcal{H}^{2}J_{0\tau}\left(\sqrt{3}\phi_{1\tau}-3\phi_{1\text{i}}'\right)e^{-2\tau+ikr}\label{vcoupling1}
\end{eqnarray}
Equating now eqs. \eqref{emcvector} and \eqref{vcoupling1},
and summing over the index $i$, one may easily find
the following scalar equation 
\begin{eqnarray}
{ik\mathcal{H}^{2}e^{-2\tau+i\mathbf{k}\cdot\mathbf{r}}\left(4\pi J_{0\tau}-\mu^{2}\phi_{0\tau}\right)\left(\sqrt{3}\phi_{1\tau}-A_{1}'\right)}=0\label{count1}
\end{eqnarray}
where $A_{1}$ is $\underset{i}{\Sigma}~\phi_{1i}$ and accordingly $A_{1}'=\underset{i}{\Sigma}~\phi_{1i}'$.
However, by using the vector field equation \eqref{eoms},
one can readily conclude that the first parenthesis of (\ref{count1}) vanishes, see
equation \eqref{bakv}. Moreover, by using the field equation of $\phi_{\alpha}$, we show now that the second parentheses is also zero. Let us first take the divergence of \eqref{eoms} by keeping in mind that $\nabla_{\alpha}\phi^{\alpha}=0$
and $\nabla_{\alpha}J^{\alpha}=0$. Consequently we arrive at a constraint
identity on $B_{\alpha\beta}$, namely  $\nabla_{\alpha}\nabla_{\beta}B^{\alpha\beta}=0$.
By linearizing this constraint we find
\begin{equation}
\begin{small}
 \left(\mathcal{H}^{5}-\mathcal{H}\right)(f')+\mathcal{F}(\mathcal{H})f=0\label{m1}
\end{small}
\end{equation}
where the function $f$ is defined as $f=\sqrt{3}\phi_{1\tau}-A_{1}'$ and $\mathcal{F}$ is $\left(6e^{2\tau}\mathcal{H}^{3}-4\mathcal{H}^{4}\mathcal{H}'-e^{2\tau}\mathcal{H}^{2}\mathcal{H}'-\mathcal{H}'+4\mathcal{H}^{5}\right)$.
One may straightforwardly conclude that $f=0$. On the other hand,
spatial isotropy implies that $A=\phi_{1x}+\phi_{1y}+\phi_{1z}=3\phi_{1j}$,
where $j=1,2,3$. This directly yields a simple differential equation
between vector field components as 
\begin{equation}
\phi'_{1j}=\frac{\phi_{1\tau}}{\sqrt{3}}\label{m4}
\end{equation}
which is equivalent to $A_{1}'=\sqrt{3}\phi_{1\tau}$ and consequently
we have $\nabla_{\mu}\delta T_{\nu~(\phi_{\alpha})}^{\mu}$=0, or
equivalently $(B_{\alpha\nu}J^{\alpha})_{1}=0$. This result has an
interesting consequence. In fact, it shows that
$T_{\text{(m)}}^{\mu\nu}$ is conserved even in the linearized limit,
i.e. $\nabla_{\mu}\delta T_{\nu~(\text{m})}^{\mu}=0$, see equation
\eqref{mcoupling1}. This conservation equation along with the relation
\eqref{tem} leads to the following expressions
\begin{align}
&\hspace{-0.8cm}\delta k^{2}c_{s}^{2}+\theta\mathcal{H}\left((\omega+1)\left(\mathcal{H}(3\omega-1)-\mathcal{H}'\right)-\mathcal{H}\omega'\right)\nonumber\\
&\hspace{-0.8cm} +\mathcal{H}^{2}(\omega+1)\theta'-k^{2}\Psi(\omega+1)=0\label{emc}\\
&\hspace{-0.8cm} 3\delta c_{\text{s}}^{2}+\delta'+\theta+3\Phi'-3\delta w+\theta w+3w\Phi'=0\label{emc2}
\end{align}
where \eqref{emc} is obtained from the spatial component $\nabla_{\mu}\delta T_{\text{i}~(\text{m})}^{\mu}=0$,
and \eqref{emc2} is the corresponding time component.

Let us now summarize this section by considering the number of unknowns and equations. There are seven unknown perturbation quantities:
$G_{1}$, $\phi_{1\tau}$, $\phi_{1i}$, $\rho$, $\theta$, $\Psi$
and $\Phi$. Accordingly, we need seven equations
to describe the evolution of the perturbed quantities. Three equations are given by the conservation equations. More specifically,
the conservation equation of $T_{(\mathcal{G})}^{\mu\nu}$, i.e. equation \eqref{Coupling}
yields a differential equation for $\mathcal{G}_{1}$, see equation \eqref{approg}
in the next section, while the conservation equation for $T_{(\text{m})}^{\mu\nu}$
gives the two differential equations \eqref{emc} and \eqref{emc2}. Moreover, using the identity $\nabla_{\alpha}\nabla_{\beta}B^{\alpha\beta}=0$,
we found a relation between the components of the vector field, see equation \eqref{m4}.
Consequently, we still need three equations to construct a complete
set of equations. To find
these three equations, in the next section we use the time component
of the vector field equation \eqref{eoms}, along with the off-diagonal
and time components of the field equation \eqref{ein}.

\section{Perturbations in the sub-horizon scale}

\label{sec-3} In this section, we investigate the evolution of the density
parameter $\delta$ in the sub-horizon scale. Specifically, the sub-horizon
scale corresponds to the scale at which the physical wavelength $2\pi a/k$
 is much smaller than the Hubble radius $1/H$. In
order to apply the sub-horizon limit to the perturbed equations, we
introduce the dimensionless length parameter $\lambda=\mathcal{H}/k$ and perform the limit $\lambda\ll 1$, keeping only terms up to the lowest order.
We restrict ourselves to the matter dominated
epoch in MOG, where  structure formation occurs. Therefore it is
natural to expect that the equation of state parameter and the sound
speed are zero, i.e. $\omega=0$ and $c_{\text{s}}^{2}=0$.

Keeping these assumptions in mind, the off-diagonal component of \eqref{ein},
leads to the following relation 
\begin{equation}
\Phi+\Psi=-\frac{\mathcal{G}_{1}}{\mathcal{G}_{0}}\label{aniso}
\end{equation}
Now, equation \eqref{emc2} can be written as 
\begin{equation}
\delta'+3\,(c_{s}^{2}-\omega)\delta=-(\theta+3\Phi')(\omega+1)\label{deltap}
\end{equation}
On the other hand equation \eqref{emc} gives 
\begin{small}
\begin{equation}
\hspace{-0.6cm}\theta'-\left(\frac{6\,\omega+3\,\omega_{\text{t}}-1}{2}-\frac{\omega'}{1+\omega}\right)\theta=\frac{1}{\lambda^{2}}\left(\frac{c_{\text{s}}^{2}\delta}{1+\omega}+\Psi\right)\label{thetap}
\end{equation}
\end{small}
where we have conveniently defined the total equation of state parameter
$\omega_{\text{t}}$ as follows 
\begin{equation}
\frac{\mathcal{H'}}{\mathcal{H}}=1+\frac{H'}{H}=-\frac{1}{2}-\frac{3}{2}\,\omega_{\text{t}}.\label{hdef}
\end{equation}
Differentiating \eqref{deltap} with respect to $\tau=\text{ln}a$,
and combining with \eqref{thetap}, we arrive at
\begin{equation}
\delta''=\frac{1}{2}(3\,\omega_{\text{t}}-1)\left(\delta'+3\Phi'\right)+\frac{1}{\lambda^{2}}\Big(\frac{\mathcal{G}_{1}}{\mathcal{G}_{0}}+\Phi\Big)-3\Phi''.
\label{dppsolve}
\end{equation}
As we already mentioned, in the sub-horizon limit, we ignore $\Phi''$
and $\Phi'$ in comparison with $\frac{\Phi}{\lambda^{2}}$. In order
to find a relation between $\Phi$ and $\mathcal{G}_{1}$, we exploit the perturbed
time component of equation (\ref{ein}) and also we apply the sub-horizon limit for $\mathcal{G}$ to find
\begin{equation}
\frac{\Phi}{\lambda^{2}}+\frac{1}{\mathcal{G}_{0}}\left(\frac{\mathcal{G}_{1}}{2\lambda^{2}}-\frac{4\pi\rho\delta}{e^{-2\tau}\mathcal{H}^{2}}-\mu^{2}\phi_{0\tau}\phi_{1\tau}\right)=0\label{metrone0}
\end{equation}
Now we need to find the last term, i.e., $\mu^{2}\phi_{0\tau}\phi_{1\tau}$,
in terms of the other perturbations. In order to quantify this term,
we perturb the vector field equation \eqref{eoms}. The vector field $J^{\alpha}$ is defined \cite{Moffat:2015bdaf} as $\kappa\rho_{\text{m}}u^{\alpha}$,
where $u^{\alpha}$ is the four-velocity, and $\rho_{\text{m}}$ is
the matter density. Using equation \eqref{eoms} the vector field $J_{\alpha}$ is also specified. One can straightforwardly check that the constraints on $\phi_{\alpha}$ and $J_{\alpha}$, i.e. $\nabla_{\alpha}\phi^{\alpha}=0$
and $\nabla_{\alpha}J^{\alpha}=0$, do not add new first order equations
for the vector fields.

Now, let us return to equation (\ref{metrone0}) in which one can
replace $\mu^{2}\phi_{0\tau}\phi_{1\tau}$ using equations \eqref{eoms}
and the definition of $J^{\alpha}$, as explained above. We ignore the term including $\Psi$
in comparison with $\frac{\Phi}{\lambda^{2}}$. Finally equation (\ref{metrone0})
takes the following form
\begin{equation}
\hspace{-0.8cm}\frac{\Phi}{\lambda^{2}}+\frac{1}{\mathcal{G}_{0}}\left(\frac{\mathcal{G}_{1}}{2\lambda^{2}}-\frac{4\pi\rho\delta}{e^{-2\tau}\mathcal{H}^{2}}-\Big(\frac{4\pi\kappa\rho}{e^{-\tau}\mathcal{H}\mu}\Big)^{2}\delta\right)=0\label{metrone}
\end{equation}
In order to find the perturbed fields $\mathcal{G}_{1}$, $\Phi$ and $\Psi$, we use
the time component of the conservation equation of $T_{\mathcal{G}}^{\mu\nu}$. In fact, we use this relation instead of the field equation of $\mathcal{G}$
given by \eqref{boxg}. Keeping the lowest order of $\lambda$, the result takes the following simple form 
\begin{equation}
\frac{\mathcal{G}_{1}}{\lambda^{2}}+\frac{\mathcal{G}_{0}}{8\pi\lambda^{2}}(\Psi+2\Phi)=0\label{approg}
\end{equation}
Now, we have equations \eqref{aniso}, (\ref{metrone}) and \eqref{approg}
for three unknowns $\mathcal{G}_{1}$, $\Phi$ and $\Psi$. Some algebraic manipulations
gives 
\begin{eqnarray}
 & \Psi & =-\frac{16\pi(4\pi-1)\lambda^{2}\rho\left(4\pi\kappa^{2}\rho+\mu^{2}\right)}{(16\pi-3)\mathcal{G}\mathcal{H}^{2}e^{-2\tau}\mu^{2}}\delta\nonumber \\
 & \Phi & =\frac{8\pi(8\pi-1)\lambda^{2}\rho\left(4\pi\kappa^{2}\rho+\mu^{2}\right)}{(16\pi-3)\mathcal{G}\mathcal{H}^{2}e^{-2\tau}\mu^{2}}\delta\nonumber \\
 & \mathcal{G}_{1} & =-\frac{8\pi\lambda^{2}\rho\left(4\pi\kappa^{2}\rho+\mu^{2}\right)}{(16\pi-3)e^{-2\tau}\mathcal{H}^{2}\mu^{2}}\delta\label{allperts}
\end{eqnarray}
As an aside, one can immediately
derive the anisotropic stress $\eta=-\Phi/\Psi$ as follows
\begin{equation}
\eta=\frac{8\pi-1}{8\pi-2}\approx1.04
\end{equation}
This quantity, which is unity in the standard case, can be measured
by combining weak lensing and galaxy clustering. Although present
constraints on this parameter are still very weak, in \cite{Pinho:2018unz} it has been
shown that a Euclid-like survey can measure a constant $\eta$ to
within a few percent. This might then be an additional way to distinguish
MOG from standard gravity.

 From now on, we focus on the matter perturbation growth. In the context
of MOG, the evolution of density contrast in the matter dominated era takes the
form 
\begin{equation}
\delta''+\Big(\frac{1}{2}-\frac{3\,\omega_{\text{t}}}{2}\Big)\delta'-\frac{4\pi-1}{16\pi-3}\Big(\frac{16\pi\left(4\pi\kappa^{2}\rho+\mu^{2}\right)}{\mathcal{G}_0 \mathcal{H}^{2}\mu^{2}e^{-2\tau}}\Big)\delta\rho=0.\label{delta1}
\end{equation}
To simplify this equation, we first replace $\mathcal{G}_{0}$ by $1/G_{0}$,
then $\rho$ by $\frac{3\mathcal{H}^{2}e^{-2\tau}\Omega_{\text{m}}}{8\pi G_{0}}$
and the term $\mathcal{H}e^{-\tau}$ by $H$. Finally, we find 
\begin{small}
\begin{equation}
\delta''+\left(\frac{1}{2}-\frac{3\,\omega_{\text{t}}}{2}\right)\delta'-\frac{(4\pi-1)}{(16\pi-3)}\left(\frac{6H^{2}\kappa^{2}\Omega_{m}}{G_{0}\mu^{2}}+4\right)\frac{3}{2}\Omega_{m}\delta=0.\label{deltaf}
\end{equation}
\end{small}
The coefficient
\begin{equation}
Y\equiv\frac{(4\pi-1)}{(16\pi-3)}\left(\frac{6H^{2}\kappa^{2}\Omega_{m}}{G_{0}\mu^{2}}+4\right)
\end{equation}
represents the modification of the Poisson equation induced by MOG
terms. One has $Y=1$ in the standard gravity. This coefficient is variously
denoted as $G_{\rm eff}$ or $\mu$ in current literature. Together with
$\eta$ given above, it fully characterizes the theory at linear, quasi-static scales.

In order to simplify equation \eqref{deltaf}, we use some results from \cite{Jamali:2017zrh}
to write the following relation. Moreover, one can check that $\Omega_{\mu}$ is almost negligible during the cosmic
evolution, which confirms our assumption that $\mu$ is almost constant. 
\begin{equation}
\frac{12H^{2}\kappa^{2}\Omega_{\text{m}}^{2}}{G_{0}\mu^{2}}=1-(\Omega_{\text{m}}+\Omega_{R}+\Omega_{G}+\Omega_{\Lambda})\label{m6}
\end{equation}
and combining equations \eqref{deltaf} and \eqref{m6}, we arrive at
\begin{equation}
\delta''+\left(\frac{1}{2}-\frac{3\,\omega_{\text{t}}}{2}\right)\delta'-\frac{3}{2}\Omega_{m}Y_{\mathrm{MOG}}\delta=0.\label{deltaf2}
\end{equation}
where now we see that
\begin{equation}
Y_{\mathrm{MOG}}=\frac{(4\pi-1)}{2\Omega_{\mathrm{m}}(16\pi-3)}\Big(1-\Omega_{R}-\Omega_{G}-\Omega_{\Lambda}+7\Omega_{\text{m}}\Big)
\end{equation}

\begin{figure}
\centering \includegraphics[width=0.46\textwidth]{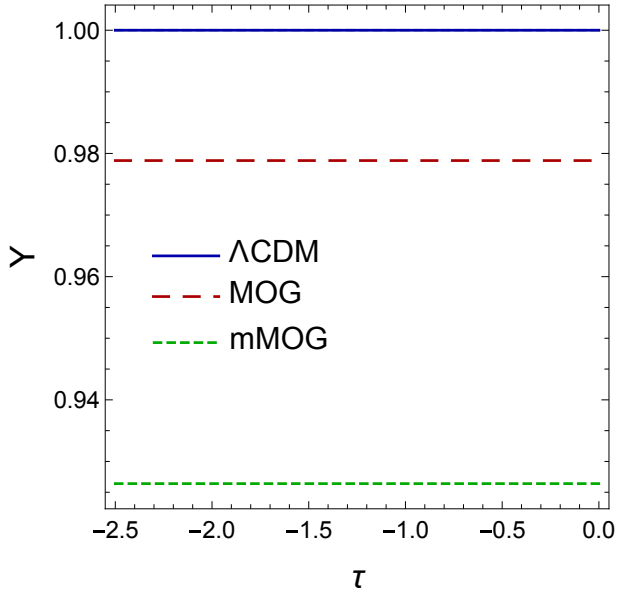} \caption{The evolution of the coefficients $Y_{\mathrm{MOG}}$ and $Y_{\mathrm{mMOG}}$ for $c=0.33 \times 8 \pi$
 is compared to the $\Lambda$CDM case, i.e., the blue line $Y=1$. The smaller $Y$ for MOG and mMOG results in a slower growth rate.}
\label{coef} 
\end{figure}
This is the main expression needed to investigate
the linear perturbations in MOG. Let us compare it with the
corresponding equation in $\Lambda$CDM, namely 
\begin{equation}
\delta''+\left(\frac{1}{2}-\frac{3\,\omega_{\text{t}}}{2}\right)\delta'-\frac{3}{2} \Omega_{\text{m}}\delta =0.\label{deltastnd}
\end{equation}
It is clear that the main difference between equations \eqref{deltaf2}
and \eqref{deltastnd} is the coefficient $Y$, since $\omega_{\text{t}}$
evolves similarly in both MOG and $\Lambda$CDM \cite{Jamali:2017zrh}.
 This coefficient can qualitatively specify whether
the growth rate in MOG is lower or higher than that of $\Lambda$CDM.
Using the numerical solutions, the evolution of $Y$ is shown in Fig. \ref{coef}, for MOG, mMOG and $\Lambda$CDM.
The magnitude of $Y$ in MOG and mMOG is smaller than $\Lambda$CDM. Therefore one may expect a slower growth rate for matter perturbations in MOG.
\section{Numerical integration}
Now, let us solve \eqref{deltaf2} by choosing a suitable set of initial conditions. We begin with the simplest choice, that is,
 the same initial conditions  of $\Lambda$CDM in the deep matter
dominated phase, namely 
\begin{equation}
\delta'(\tau^{*})=\delta(\tau^{*}),~~~~~~~\delta(\tau^{*})=a^{*}.
\label{m10}
\end{equation}
where the initial $\tau^{*}=\ln a^{*}$ is taken at $\tau=-2.5$,
which corresponds to the redshift $z^{*}\simeq11$. As advertised, and as we 
found, solving \eqref{deltaf2}, $\delta$ in MOG grows slower than in $\Lambda$CDM. 

The initial conditions
in modified theories of gravity, in principle, can be different from
$\Lambda$CDM, for example see \cite{DiPorto:2007ovd}. Consequently,
we generalize the initial conditions as
\begin{equation}
\delta'(\tau^{*})=\beta\,\delta(\tau^{*})\label{m8}
\end{equation}
where the new parameter, $\beta$, expresses the deviation from $\Lambda$CDM initial conditions.
We need to consider only $\beta<1$ since we checked analytically that in matter dominated era of MOG, we have always $\delta'/\delta<1$. We also checked that for any $\beta$ the evolution of $\delta$ reverts soon
back to the case $\beta=1$. The conclusion that the growth
of $\delta$ in MOG is slower than $\Lambda$CDM does not  depend therefore on the initial conditions.

It is also instructive to investigate the growth rate  $f$
of matter perturbations, defined as follows 
\begin{equation}
f=\frac{\delta'}{\delta}=\frac{d\ln\delta}{d\ln a}\label{fdef}
\end{equation}
Equation \eqref{deltaf2} can be written in terms of $f$ as 
\begin{equation}
f'+f^{2}+\frac{1}{2}(1-3\,\omega_{\text{t}})f-\frac{3}{2}\Omega_{\mathrm{m}}Y_{\mathrm{MOG}}=0.\label{grow1}
\end{equation}
To solve this equation, we need only one initial condition, \eqref{m8}, namely $f(z^{*})=\beta$.
\begin{figure}
\centering
  \includegraphics[width=0.46\textwidth]{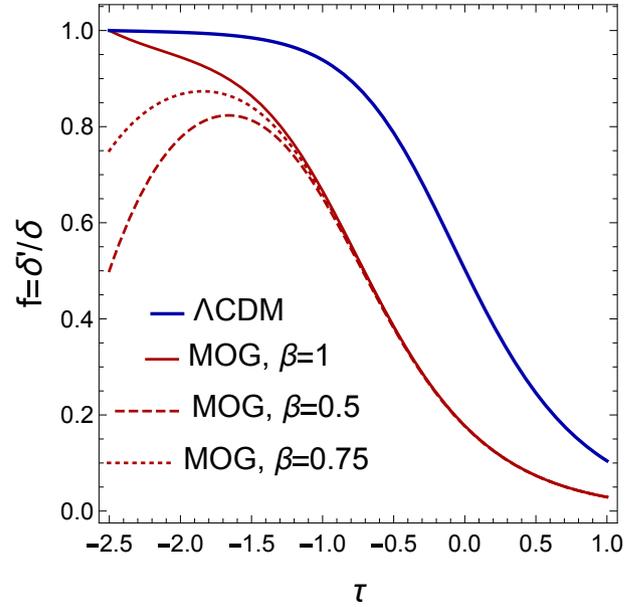}
\caption{
  The growth
rate function in MOG and $\Lambda$CDM. The initial conditions are $f(-2.5)=1$ for $\Lambda$CDM and $f(-2.5)=\beta$ for MOG. }
\label{mldelta} 
\end{figure}
The result is illustrated in  Fig. \ref{mldelta}, that clearly confirms the growth parameter $f$ in MOG is
always smaller than the standard case, regardless of $\beta$. We now proceed  to evaluate the growth rate in mMOG,
a modified version of MOG in which a new free parameter $c$ has been incorporated by changing the kinetic
energy contribution of the scalar field $G$ and it turns out that for the value  $c=0.33 \times 8 \pi$, it brings the sound horizon  angular size  compatible with the observation, while at the same time achieving a viable sequence of cosmological epochs \cite{Jamali:2017zrh}.  Following the same steps as for MOG, we find
\begin{equation}
\delta''+\left(\frac{1}{2}-\frac{3\,\omega_{\text{t}}}{2}\right)\delta'-\frac{3}{2}Y_{\mathrm{mMOG}}\Omega_{m}\delta=0.\label{deltammog}
\end{equation}
where 
\begin{equation}
Y_{\mathrm{mMOG}}=\frac{(c-2)}{(8c-12)\Omega_{m}}\Big(1-\Omega_{R}-\Omega_{G}-\Omega_{\Lambda}+7\Omega_{m}\Big)
\end{equation}
where $\Omega_{G}=\frac{G'}{G}-\frac{c}{6}(\frac{G'}{G})^{2}$
and the other $\Omega$'s are the same as in MOG. 
In order
to avoid the existence of tachyonic instability, we  restrict
ourselves to positive $c$. 
\begin{figure}
\centering 
 \includegraphics[width=0.49\textwidth]{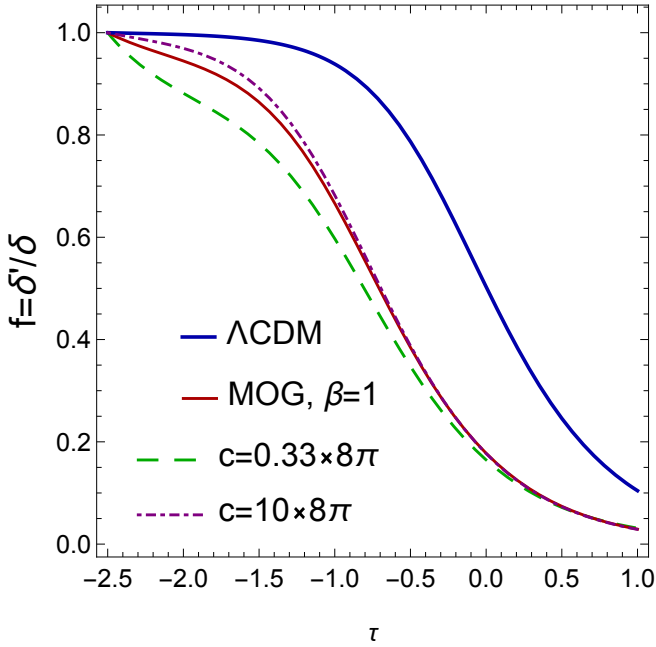}
\caption{\label{Range}
  The behavior of $f$
in mMOG for different choices of $c$ and $\Lambda$CDM. The growth of perturbations in mMOG is slower than $\Lambda$CDM, using the same initial condition, $f(-2.5)=1$}
\label{dfmmog} 
\end{figure} 
We find the exact numeric
solution of \eqref{deltammog} for the same initial conditions
described for MOG. Our solution shows that the new parameter $c$ does not lead to a significant deviation from MOG.
The corresponding growth rate parameter $f$, for different values of $c$, is plotted in Fig. \ref{dfmmog}. 

We can simply conclude that
both MOG and mMOG lead to slower matter growth compared to $\Lambda$CDM. The reason is that the rate of growth depends heavily on the fraction of matter compared
to other energy contributions. If the other homogeneous
fields decrease this fraction, as it occurs here, the effect  overcompensates the
increase of gravitational strength, and the result is less growth.

In the following section, we compare our results with the relevant
observation and discuss the viability of MOG as an alternative to
dark matter particles. 

\section{Comparison with observation}

\label{sec-4} In the previous section, we have obtained the evolution
of the growth function $f$
in the context of MOG and a slightly modified version called mMOG.
Here, we use the available data for $f\sigma_{8}$
to compare
\eqref{deltammog} with observation. The RSD parameter, $f\sigma_{8}$, is
defined as 
\begin{equation}
f\sigma_{8}\equiv\sigma_{8}(z)\frac{\delta'(z)}{\delta(z)}\label{fsigma}; \quad \text{where} \quad \sigma_{8}(z)=\sigma_{8}^{0}~\frac{\delta(z)}{\delta(0)}.
\end{equation}

To obtain the current value of $\sigma_{8}$, it is necessary to specify
the underlying gravity theory. Then $\sigma_{8}^{0}$ is determined
through model dependent observations such as CMB power spectrum \cite{Ade:2015xua},
weak lensing \cite{More:2014uva}
and abundance of clusters \cite{Ade:2015fva}.
Consequently, $\sigma_{8}^{0}$ is a model-dependent quantity \cite{Nersisyan:2017mgj},
and naturally one may expect a different value for it in MOG compared
with $\Lambda$CDM (where  $\sigma_{8}^{0}=0.802\pm0.018$
\cite{Ade:2015xua}), as we are going to find now.

\begin{figure}
 \includegraphics[width=0.46\textwidth]{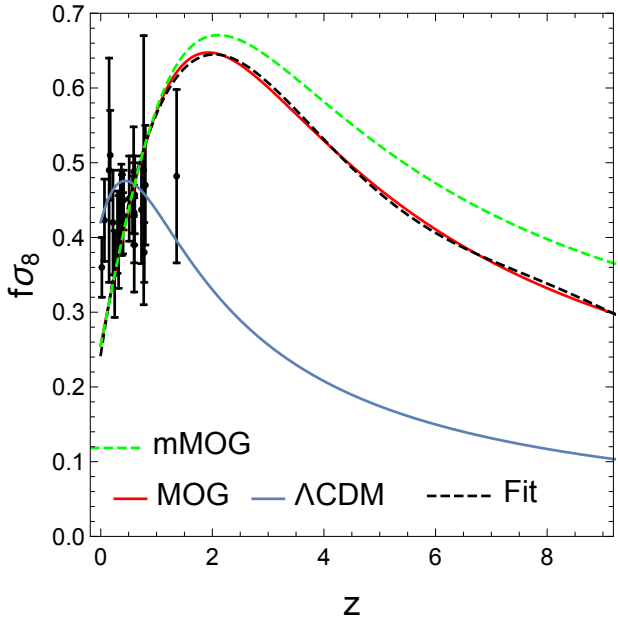}~~~~~~
\caption{
 The long term evolution of $f\sigma_{8}^{0}$ for $\Lambda$CDM, mMOG with $c=0.33 \times 8 \pi$,
MOG and the best fit with the 5th order polynomial $\sum_{k}A_{k}a^{k}$,
where $A_{5}=8\times10^{-5}$, $A_{4}=-0.00275$, $A_{3}=0.03544$,
$A_{2}=-0.20970$, $A_{1}=0.49936$ and $A_{0}=0.24289$.
}
\label{mlfsigma} 
\end{figure}

We assume that $f\sigma_{8}(z)$ data are also valid in MOG\footnote{It is necessary to mention that also  $f\sigma_{8}$ is not completely
model independent and in principle, one has to find it for the model
under consideration. In the case of MOG, we can use the available
data points, since  MOG is designed to recover the same evolution
for the background quantities as in $\Lambda$CDM}. Then we solve the perturbation equation \eqref{deltammog}, and
by fitting to $f\sigma_{8}(z)$ data, we predict the best value for
$\sigma_{8}^{0}$ in MOG. It is worth mentioning that the available
data for $f\sigma_{8}$ lie in the redshift range $0\le z\le1.2$
(see Fig. \ref{mlfsigma}). In this interval, the baryonic matter,
the cosmological constant and the scalar field $G$ have non-zero
contribution to the energy budget of the Universe, see Fig. 1 in \cite{Jamali:2017zrh}.
On the other hand, one can ignore radiation in equation \eqref{deltammog}.

As we already discussed, equation \eqref{deltaf2} needs two initial
conditions to be solved. Consequently, we have two free parameters
$\sigma_{8}^{0}$ and $\beta$. However, we have already discussed in the
previous section that the evolution of perturbations does
not change significantly with initial conditions, essentially because  the growing mode dominates, regardless
of the initial conditions. Therefore, without loss of generality,
we set $\beta=1$ as in $\Lambda$CDM.

For the data points, $\mathcal{D}_{i}$, we used
Table II in \cite{Albarran:2016mdu}. In the case of independent data
points, the likelihood function $\mathcal{L}$ is given by a simple
relation, 
\begin{equation}
\mathcal{L}=A~\text{exp}\big[-\chi^{2}/2\big]
\end{equation}
in which $A$ is a normalization constant and $\chi^{2}$ is defined
as 
\begin{equation}
\chi^{2}=\sum_{i}\frac{(\mathcal{D}_{i}-\mathcal{T}_{i})^{2}}{\sigma_{i}^{2}}\label{chi}
\end{equation}
where $\mathcal{D}_{i}$ and $\mathcal{T}_{i}$ refer to the predicted
value of an observable by data and theory, respectively. Furthermore,
$\sigma_{i}$ is the error associated with the $i$th data point. Specifically,
in our case, we have 
\begin{equation}
\mathcal{L}=\sum_{i}~A~\text{exp}\bigg[-\frac{1}{2}\Big(\frac{\mathcal{D}_{i}(z)-\sigma_{8}^{0}\times\mathcal{T}_{i}(z)}{\sigma_{i}}\Big)^{2}\bigg].
\end{equation}
We performed the likelihood analysis and found the maximum
of $\sigma_{8}^{0}$ are $1.44$ and $1.59$, for MOG and
mMOG, respectively. Since in MOG matter is entirely constituted by the baryonic fraction,  we anticipated a larger value for $\sigma_{8}^{0}$, compared to $\Lambda$CDM, to compensate for the smaller matter gravitational pull. A similar situation can be seen in \cite{MacCrann:2014wfa}.

 In Fig. \ref{mlfsigma}, we plot $f\sigma_{8}$ for MOG, mMOG
and $\Lambda$CDM along with the data points. We have shown the long-term
evolution of $f\sigma_{8}$ for all the models, to have a better insight about the models. We also fit a polynomial to the curves in
MOG and mMOG.
 In the case of $\Lambda$CDM we picked the reported
$\sigma_{8}^{0}$ in \cite{Ade:2015xua}, while for MOG and mMOG we
used the result of our likelihood analysis. As the plot clearly shows,
the evolution of $f\sigma_{8}$ in MOG and $\Lambda$CDM is significantly
different. Although $\chi^{2}/$dof is smaller in MOG, one needs more
data points to decide which model can fit the data more accurately.

We summarize all the results obtained from the likelihood analysis
in Table \ref{tab1}. The main conclusion is that both MOG and mMOG predict larger
values for $\sigma_{8}^{0}$. Of course, to make a reliable decision
on the viability of MOG as an alternative theory of dark matter, it
is necessary to measure $\sigma_{8}^{0}$ from other relevant observations,
like CMB and lensing \cite{Hildebrandt:2016iqg}. We leave this point to future studies. 
In TeVeS, a relativistic version for Modified Newtonian
Dynamics (MOND) \cite{Bekenstein:2004ne}, the vector field
can play a role similar to cold dark matter \cite{Dodelson:2006zt}
and increase the matter growth rate. However, our analysis shows that
this is not the case in MOG, and the Proca vector field does not expedite
the perturbation growth. Therefore, our conclusion can be considered
a challenge to the viability of this theory.

It is necessary to mention another important issue. It has been shown in \cite{Dodelson:2011qv} that TeVeS leads to a huge enhancement of baryonic acoustic oscillations (BAO). Such behavior raises a serious challenge for the viability of TeVeS and is inconsistent with the observations. A similar enhancement in BAO can be seen in the context of MOG reported in \cite{Moffat:2011rp}. The rapid oscillations in the power spectrum may be simply related to the fact that in absence of dark matter, baryons weigh more. However, it is necessary to revisit this important issue in MOG. The results presented in \cite{Moffat:2011rp} are based on several approximations and analytic descriptions. On the other hand, the conventional and more reliable way to derive the angular power spectrum of CMB is to use the relevant numeric codes like CAMB \cite{Lewis:2002ah} and explore the full set of parameters. Therefore, this code should be modified to include MOG effects. Of course, this is not easy and is well beyond the scope of our paper.

\section{Discussion and Conclusion}

\label{sec-5} In this paper, we investigated the cosmological perturbations
in the context of a Scalar-Tensor-Vector theory of gravity known as
MOG. As in the standard case, we started from the
modified Friedmann equations and introduced the perturbed metric in
the Newtonian gauge. We assumed that the matter content of the universe
is a perfect fluid and, without imposing restrictive assumptions on
the evolution of the fields, we found the first order perturbed field
equations.

It is well known in the literature that any deviation from $\Lambda$CDM
in matter dominated era may substantially influence the structure
formation scenario. In order to consider this in detail, we evaluated the evolution
of matter perturbations, $\delta$, and the growth function, $f=\delta'/\delta$,
in the context of MOG. Since the growth of gravitational seeds starts
in the sub-horizon scale, we have considered the perturbed equations
in the sub-horizon limit in the matter dominated epoch.

 We also presented
a similar description for mMOG, which is a different version of MOG
compatible with the sound horizon observations. Our main result is   that the growth of matter perturbations in both MOG and mMOG is slower than in $\Lambda$CDM.
In fact, this is a surprising result, since in all the modified gravity
theories aiming at replacing dark matter, the gravitational force should be strengthened in
the weak field limit, in order to explain the flat galactic rotation curves, and other observations, without the pull of dark matter. However, we have shown that the reduced matter content of MOG overcompensates the extra gravitational force.

\begin{table}[width=.9\linewidth,cols=3,pos=h]
\caption{$\sigma_{8}^{0}$ and $\chi^{2}$/dof for $\Lambda$CDM, MOG and mMOG. }
\label{tab1} 
\begin{tabular*}{\tblwidth}{@{} LLLL@{} }
\toprule
$\Lambda$CDM & MOG & mMOG \\
\midrule
$\sigma_{8}^{0}$=0.82 \cite{Ade:2015xua} & $\sigma_{8}^{0}=1.44$ & $\sigma_{8}^{0}=1.59$ \\
$\chi^{2}$/dof=$0.703$ & $\chi^{2}$/dof=$0.651$ & $\chi^{2}$/dof=$0.647$ \\
\bottomrule
\end{tabular*}
\end{table}

We wrote down the full set of perturbation equations and determined
the two modified gravity parameters, $\eta$ and $Y$. We then compared
MOG, mMOG and $\Lambda$CDM with the observed $f\sigma_{8}$, and
found that MOG and mMOG require higher values for $\sigma_{8}^{0}$.
The RSD data do not yet rule out MOG but the high value of $\sigma_{8}$
seems problematic when compared to recent estimates due to lensing. Therefore we conclude that although MOG is not yet ruled out,
a full analysis of CMB and lensing data  will
provide a strong challenge to MOG.

\section*{Acknowledgement}
Sara Jamali thanks Henrik Nersisyan and Malihe Siavoshan
for useful discussions. She also would like to thank the Institute
for Theoretical Physics, University of Heidelberg for a very kind
hospitality, during which some parts of this work have been done.
Mahmood Roshan would like to thank Sohrab Rahvar for useful discussions.

\printcredits

\bibliographystyle{unsrt}

\end{document}